\documentclass[pra,aps,twocolumn,showpacs]{revtex4}
\usepackage{graphicx}

\def\norc #1 {\ {\Vert} #1 {\Vert_{c}}}
\def\norN #1 {\ {\Vert} #1 {\Vert_{N}}}
\def\nor #1 {\ {\Vert} #1 {\Vert}}

\def \Ninf#1 {  {\Vert {#1} \Vert_\infty } }
\def \NLip#1 {  {\Vert {#1} \Vert_\theta } }
\def \Norm#1 {  {\Vert {#1} \Vert } }

\def \bra{{ \langle}}
\def \ket{{ \rangle}}

\renewcommand{\Re}{\mathrm{\,Re}}

\newcommand{\beq}{\begin{equation}}
\newcommand{\eeq}{\end{equation}}
\newcommand{\barr}{\begin{eqnarray}}
\newcommand{\earr}{\end{eqnarray}}

\begin{document}

\def\figwidth{8.5cm}


\title{Dynamically localized systems: entanglement 
exponential sensitivity and efficient quantum simulations.}

\author{Simone Montangero}
\email{monta@sns.it}
\homepage{www.sns.it/~montangero}
\affiliation{NEST-INFM $\&$ Scuola Normale Superiore, Piazza dei Cavalieri 7,
56126 Pisa, Italy}

\begin{abstract}
We study the pairwise entanglement present in a quantum computer 
that simulates a dynamically localized system. We show that the
concurrence is exponentially sensitive to changes in the Hamiltonian
of the simulated system. Moreover, concurrence is exponentially
sensitive to the ``logic'' position of the qubits chosen.
These sensitivities could be experimentally checked efficiently by means 
of quantum simulations with less than ten qubits.  
We also show that the feasibility of efficient quantum simulations 
is deeply connected to the dynamical regime of the simulated system.
\end{abstract}

\date{\today}


\maketitle
The study of the entanglement in quantum chaotic systems is object 
of growing interests \cite{studyent}. These works are
based on classical simulations of quantum systems. Developing quantum 
computers will enhance our investigations power giving the possibility 
of performing nowadays unaccessible simulations. 
However, being able to perform exact quantum computation is not 
a sufficient condition to assure
that quantum simulations will be useful: 
extracting useful information in an efficient way is a difficult task
that, in some cases,  drastically decreases the efficiency
of quantum algorithms \cite{diffqc, proc}. 
Moreover, in \cite{vidal} it has been shown that 
the entanglement present in a quantum system plays an important 
role in the possibility of performing efficient classical simulations
of quantum systems. Thus, the presence of entanglement and the processes of 
quantum measurement may determine the efficiency of
quantum algorithms.\\ 
It is well known that quantum chaotic systems display exponential
sensitivity to small changes of the Hamiltonian \cite{haake} in
dynamical quantities as the fidelity \cite{fidelity}. 
In \cite{entev} it has been shown that the entanglement evolution is 
influenced by the presence of quantum chaos in a system following the 
characteristic decays of the fidelity. 
This result has been found in a system with no classical 
analogue. Here, we study the concurrence evolution between the 
qubits in a quantum computer that is running a quantum algorithm to simulate a 
quantum chaotic system, in the regime where dynamical localization occurs. 
Dynamical localization is another signature of quantum chaos: 
it has been first predicted in \cite{casati}, 
and experimentally observed in cold atoms
\cite{expqkr}. It has been related to Anderson localization 
\cite{anderson} of electrons in disordered crystals.  
Then, it is interesting to perform simulation on many-body or 
classically chaotic quantum system to check whenever
they localize or not: The analysis of localization length as a
function  of the parameters of the system is important to study 
transport phenomena. In \cite{proc} it has been shown that 
localization length can be extracted with polynomial gain 
if an  efficient quantum algorithm to simulate the system exists. 
This is not a trivial property, as the efficient extracion of 
many interesting quantities is precluded due to the exponential 
number of measurements needed to know all the wave function
coefficients.\\ 
In this paper we show that the presence of an exponentially localized wave
function implies exponentially sensitivity in the concurrence between
qubits. This sensitivity is both with respect to small changes 
in the Hamiltonian and to the pair of qubits chosen. 
The pairwise entanglement can be extracted efficiently from any 
quantum algorithm that simulates a quantum system efficiently. 
Indeed, the measure of the reduced density matrix of two qubits is 
an efficient process that does not scale with the size of the system.
Thus, both localization length and concurrence can be measured
efficiently by means of quantum simulations.  We also study the
entanglement of blocks of qubits with respect of the other and we show 
that these properties are strictly related to the
localization length. Thus,  following \cite{vidal},
the localization length determine the 
feasibility of efficient classical simulation of the system.

\begin{figure}
\centerline{\includegraphics[height=6cm]{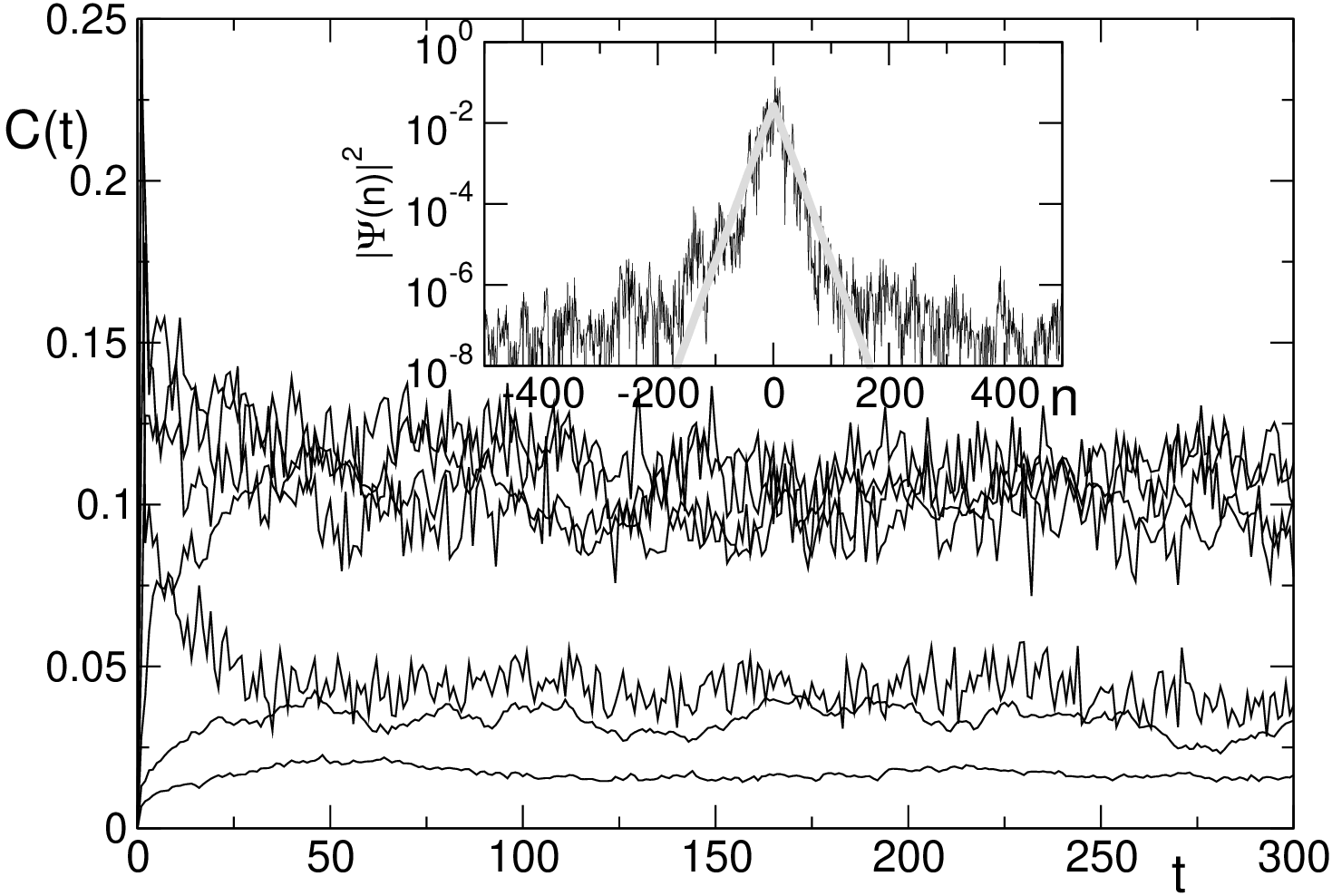}}
\caption{Concurrence of qubits $1$ and $3$ as a function of time for 
different map parameters $K=\sqrt 2$, $k=K/T$, $T=2 \pi M / 2^{n_q}$,
$n_q=10$. 
From bottom to top $M =10^4, 5 \cdot 10^3, 3 \cdot 10^2, 2 \cdot 10^3,
5 \cdot 10^2, 10^3, 7 \cdot 10^2$ and $\ell =
0.03,0.1,31,0.7,11,2.8,5.7$ respectively. Inset: peak of an exponentially
localized wave function. The grey thick line follows the law (\ref{locpsi}).} 
\label{fig1}
\end{figure}

The Quantum Sawtooth Map (QSM) is a suitable model of quantum 
chaotic system that displays dynamical localization which can be
efficently simulated with a quantum computer. 
The QSM belongs to the kicked map family and it is defined by the Hamiltonian
\begin{equation}
\hat H(t) = \frac{\hat n^2}{2} + k \frac{(\hat \theta -\pi)^2}{2}
\sum_r \delta(t -
rT),
\label{hamil}
\end{equation}
where $\delta_T(t)$ is a delta of Dirac and $r \in \Re$, $\hat n$ and 
$\hat \theta$ are canonical conjugate variables such that 
$[\hat \theta,\hat n]= \imath T$ (we set $\hbar=1$, thus
$T=\hbar_{eff}$). 
Correspondingly, the quantum dynamics is described by the Floquet operator
\begin{equation}
U = e^{\imath k (\hat \theta -\pi)^2} e^{-\imath T \hat n^2/2},
\label{qsm}
\end{equation}
where $0 < \theta < 2 \pi$ and $-N/2 < n \le N/2$. 
The wave function time evolution is computed by repeated applications of
operator (\ref{qsm}) to the initial wave function, {\it i.e.} 
$|\psi(m) \ket = U^m |\psi(0)\ket$.
The dynamics of the QSM is governed by two parameters $k,T$, while 
the dynamics of the classical correspondent map depends only
on the classical parameter $K= kT$ and displays chaotic dynamics for 
$K > 0$ and $K < -4$ \cite{sawtooth}. 
For $-4 < K < 0$ the map is described by mixed phase space 
(both integrable and chaotic).
The classical limit of map (\ref{qsm}) is recovered for $k \to \infty$
and $T \to 0$ keeping $K$ constant. 
For $T \lesssim 1$ and $K > 1$ unique quantum feature appear and QSM displays
dynamical localization. The classical chaotic diffusion in momentum is
suppressed  by quantum interference and the wave function is described
by an exponentially localized wavefunction of the form 
\begin{equation}
|\psi (n)| \approx e^{ - |n|/\ell} /\sqrt{\ell}. 
\label{locpsi}
\end{equation}
The inset of Fig.\ref{fig1} shows such typical localized wave function.
Any further evolution of the wave function is suppressed excepted from quantum
fluctuations. The QSM in the localized regime is described by the time
independent wave function (\ref{locpsi}) with 
$\ell \approx \pi^2 k^2/3$. Dynamical localization occurs after time 
$t^* \sim \ell$ \cite{simone3}. We stress the fact that the expression 
(\ref{locpsi}) is a typical signature of quantum chaos. Notice that
dynamical localization can be seen only if $\ell << N$: In the
thermodynamic limit dynamical localization always occurs.

In \cite{simone1} an efficient quantum algorithm has been presented
to simulate the map (\ref{qsm}). The algorithm is
based on the quantum Fourier transform and it exploits the exponential
efficiency of the quantum Fourier transform with respect to the classical
fast Fourier algorithm \cite{ekert}. 
If performed on a $n_q$-qubits quantum computer, it needs $O(n_q^2)$ 
elementary quantum gates, while a classical computer needs
$O(N=2^{n_q})$ elementary gates. Moreover, the QSM algorithm has 
no need of ancillary qubits: it
is already possible to see dynamical localization with less 
than ten qubits. Notice that any quantum algorithm to simulate a
quantum system is based on the introduction of the coding of the
dynamical variable in binary representation. In our case, the coding
is defined by $n = \sum_i \alpha_i 2^i$, where $n$ are the
eigenvalues of $\hat n$ describing the QSM and $\alpha_i=0,1$
depending on which level of the $i-th$ qubit is populated. 
Thus, the qubits are ordered as they represent different logical 
values, from the least significant qubit ($i=1$) to the most
significant one ($i=n_q$). 
Each of them naturally introduce a coarse graining of 
the space of the $n$ \cite{dima}.
From now on, we refer to the difference of coarse graining level 
$\Delta(i,j)=2^i-2^j$ as coarse graining distance between the qubits. 
Moreover, for the sake of simplicity, we will identify the
logic label $i$ with the spatial position. 
However, our results are general as they 
can be recovered with an appropriate mapping between spatial 
and logic position. 
 \begin{figure} 
\centerline{\includegraphics[height=6cm]{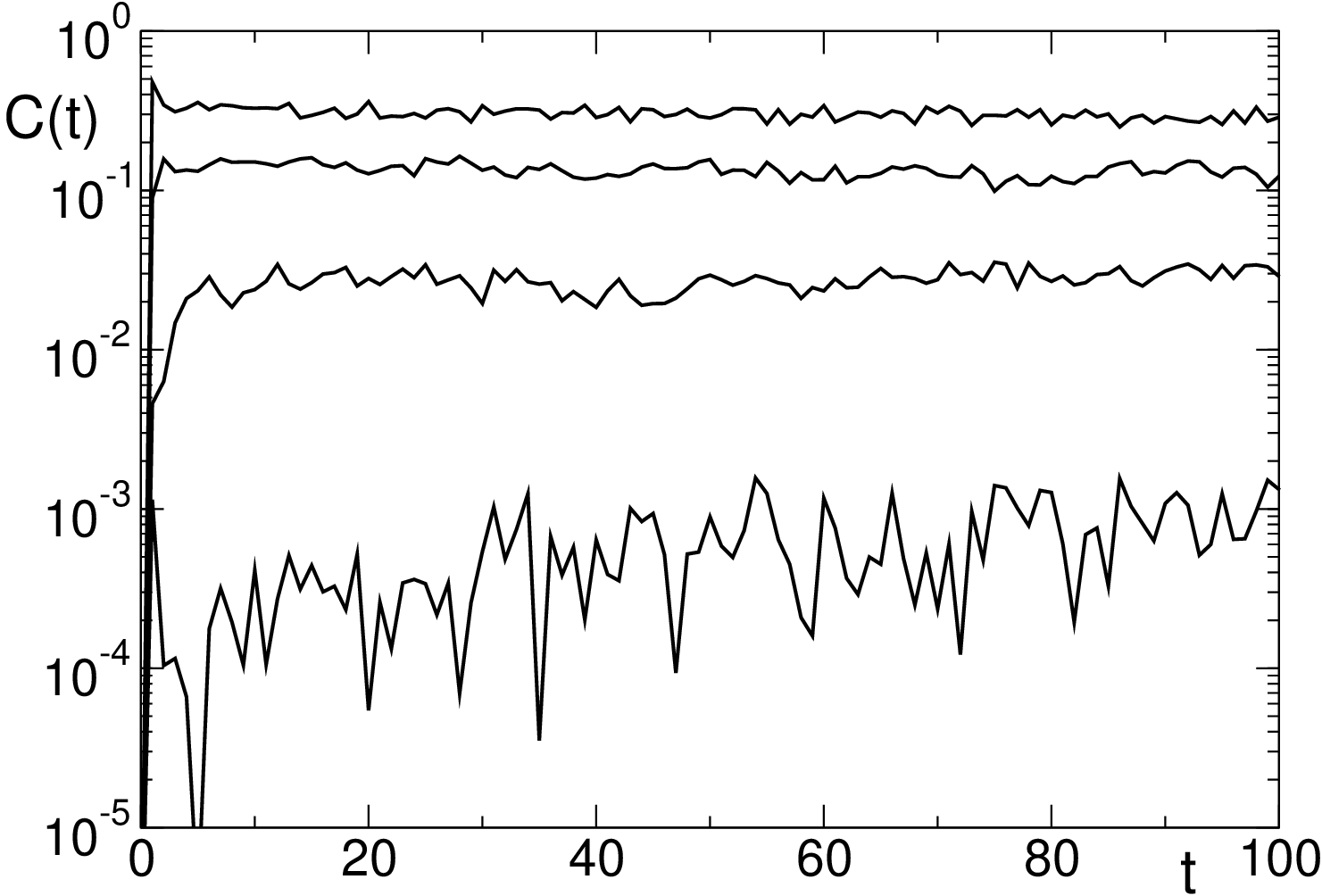}}
\caption{Concurrence of qubits $1$ and $j$ as a function of time for 
different map parameters $K=\sqrt 2$, $k=K/T$, $T=2 \pi M / 2^{n_q}$,
$n_q=10$, $M=800$, $\ell=4.36$. From top to bottom $j=2,3,4,5$.}
\label{fig2}       
\end{figure} 
It is common wisdom that entanglement is a necessary resource to
exploit the quantum computational gain. It is then natural to study
the pairwise entanglement in the qubits that describe a dynamically
localized system. We show that pairwise entanglement is present
between the qubits and  that its value depends by 
the natural ordering introduced by the 
quantum algorithm. Indeed, differently from other studies, 
the coding $n = \sum_i \alpha_i 2^i$, introduce a hyerarchy that was
not present in spin systems as in \cite{studyent}.
We show that in the regime of dynamical localization the concurrence 
value exponentially depends on the coding position of the pair of 
qubits and that it is exponentially sensitive of small change of 
the kick strength $k$. 

\begin{figure} 
\centerline{\includegraphics[height=6cm]{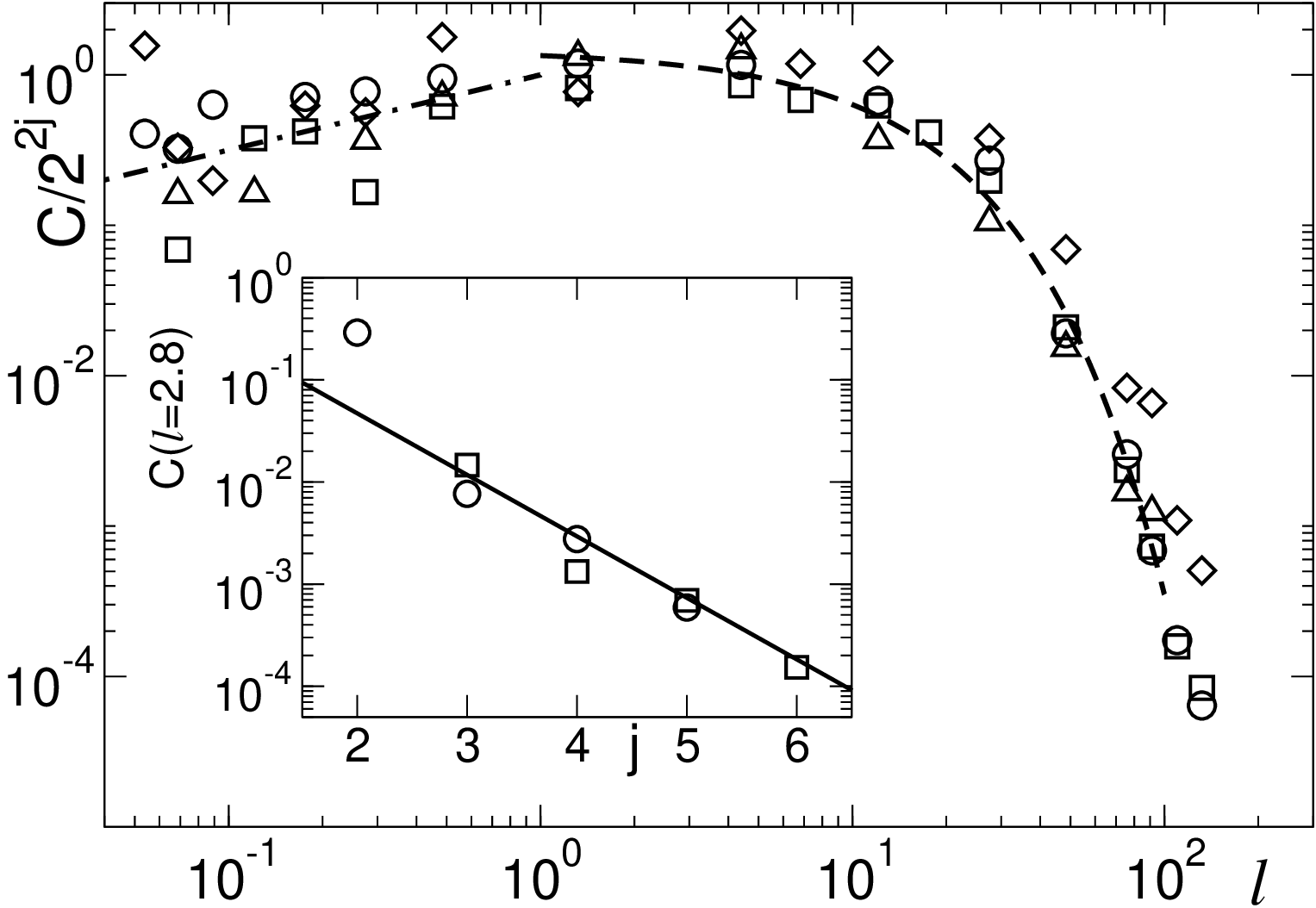}}
\caption{Concurrence saturation values between qubits $i=1$ and $j$
  as a function of localization length $\ell$. 
  Different simbols represents different qubits couples: $j=2$
  (circles), $j=3$ (squares), $j=4$ (diamonds), $j=5$ (triangles). 
  Dashed lines follows the predicted behavior $C(\ell) \sim \ell^{0.5}$
  and $C(\ell) \sim \mathrm{exp}(-A \ell)$.
  Inset: Concurence values for $\ell=2.8$ and different qubits
  couples: $i=1$ (circles) and $j-i=1$ (squares). 
  Full line represents $C(j) \sim 2^{-2j}$.}
\label{fig3}       
\end{figure} 

We quantify the pairwise entanglement present between any pair $i,j$ 
(with $i < j$) of qubits by means of concurrence \cite{conc}, defined as
$C=\max\{\lambda_1-\lambda_2-\lambda_3-\lambda_4,0\}$,
where the $\lambda_i$'s are the square roots of the eigenvalues
of the matrix $R=\rho^{i,j}\tilde{\rho}^{i,j}$, in descending
order; the spin-flipped density matrix is defined by
$\tilde{\rho}^{i,j}=(\sigma_i^y\otimes\sigma_j^y)
(\rho^{i,j})^\star (\sigma_i^y\otimes \sigma_j^y)$ (in
this definition the two qubits basis $|\alpha_i\alpha_j\rangle$ must be used).
The reduced density matrix $\rho^{i,j}$ is given by
\begin{eqnarray}
\rho^{i,j}_{l,m}(t) = 
\prod_{k=1, k\ne i,j}^{n_q} \sum_{\alpha_k = 0}^{1}
A^*_{\alpha_1 \dots \beta_l \dots \beta_m \dots \alpha_{n_q}} 
A_{\alpha_1 \dots \beta_l \dots \beta_m \dots \alpha_{n_q}}
\label{rho}
\end{eqnarray} 
where $l,m=0,1 \dots 3$, the $\beta_k$ are the binary code of $l$ and
$m$ and $A_{\alpha_1 \dots \beta_l \dots \beta_m \dots \alpha_{n_q}}=
\bra \alpha_1 \dots \beta_l \dots \beta_m \dots \alpha_{n_q} | \psi(t) \ket$.

We start with an initial wave function  equal to a momentum eigenstate 
$|\psi (t=0) \rangle = | n \ket$, we compute the
time evolution at a given time $t$ by repeated application of the
Floquet operator (\ref{qsm}) and at each time we compute the
concurrence by means of the reduced density matrix (\ref{rho}). 
We plot in Fig.\ref{fig1} the concurrence evolution of two qubits 
as a function of time for different $k$ values and fixed $K$: this
choose of the parameter values corresponds to the same classical dynamics
but different quantum dynamics with different localization time and
length.
The stationary regime is reached for larger times as $\ell$ value 
is increased. In Fig. \ref{fig2}
we plot the concurrence evolution for fixed $\ell$ and different qubit 
pairs. The different saturation levels and times are visible. 
Notice that the concurrence saturation level is not a monotonic function
of the localization length: indeed the two limiting situations, 
a momentum eigenstate and a random superposition of all the 
momentum eigenstate i.e. an ergodic wave function, 
have zero concurrence values for any qubit pair 
\cite{entev}, thus a maximally entangled case should lie in between.

In Fig. \ref{fig3} we show in details the saturation value of the 
concurrence between the less significant bit ($i=1$) and the others 
($j=2,3 \dots$)  as a function of the localization lenght. 
There are two distinct regimes: for $\ell \lesssim 1$ the
concurrence increases as $C(\ell) \sim \sqrt{\ell}$, while for $\ell
\gg 1$ the concurrence is exponential sensitive to the localization
length. The squared root dependence can be understood as follows:
The number of addend different from zero in (\ref{rho}) can be
estimated as $N_\ell= \ell/\mathcal{G} =\ell/(\Delta ~ 2^j)$.
Indeed, the number of pairs of coefficients 
$A_{\alpha_1 \dots \beta_l \dots \beta_m \dots \alpha_{n_q}}$ of coarse
graining distance $\Delta(i,j)$ in an ensemble of $\ell$ 
coefficients different from zero drops linearly with $\Delta(i,j)$, 
thus exponentially with $j$. The structure is repeated each $2^j$, 
giving the scaling $\mathcal{G} \sim \Delta ~ 2^j$.
For $\ell \lesssim 1$ only few levels of the wave function are significantly
populated, and they are ``near'', i.e. the binary code of their
position $n$ in the wave function ($\beta_1 \alpha_2 \dots \dots \beta_j 
\dots \alpha_{n_q}$) differs only for the value of
$\beta_1$. 
To estimate the sum (\ref{rho}) we aproximate the wave function with
the expression
\begin{equation}
|\psi \ket \approx \sum_{i=\bar m}^{\bar m +\ell} 
\frac{e^{\imath \phi_i}}{\sqrt{\ell}}  |i\ket,
\label{flat}
\end{equation}
where $\phi_i$ is a random phase and $\bar m + \ell/2$ the position of
the peak of the wave function \cite{simone3}. Thus the density matrix 
will be composed by diagonal terms porportional to $\rho^{i,j}_{k,k}
\sim 1/\mathcal{G}$. Notice that either $\rho^{i,j}_{0,0}$ or 
$\rho^{i,j}_{3,3}$ will be negligible as they have a coarse graining
distance is $\Delta >> \ell$. Equivalently, the off diagonal matrix
elements which differ significantly from zero are those who are composed  by
products of terms which have the smaller coarse graining distance, 
that is, the first lower and upper diagonal. 
Furthermore, if $\rho^{i,j}_{3,3}$ is negligible also
$\rho^{i,j}_{2,3}$ may be approximate by zero. 
We then compute the $R_p^{i,j}$ matrix, which mix up the density matrix
elements, leading to an approximate two by two matrix of the form:
\begin{equation}
R_p \sim 
\left(
\begin{array}{cccc}
0 & 0 & 0 & 0 \\
0 &\frac{1+\ell}{\mathcal{G}^2} & \frac{\ell}{\mathcal{G}^2}  & 0  \\
0 & \frac{\ell}{\mathcal{G}^2} & \frac{1+\ell}{\mathcal{G}^2}   & 0  \\
0 & 0 & 0 & 0
\end{array}
\right).
\label{rpert}
\end{equation}
The squared root of the non-zero eigenvalue of $R_p$ matrix is then 
proportional to $\sqrt{\ell}/\mathcal{G} \sim \sqrt{\ell}/2^{2j}$. 
Equation (\ref{rpert}) implies that the saturation value of the 
concurrence for small $\ell$ scales as
$C \sim \sqrt{\ell}/2^{2j}$: This prediction is in agreement with the
numerical data presented in Fig. \ref{fig3}.

For $\ell >> 1$ the concurrence has a completely different behavior
(Fig. \ref{fig3}):  the number of coefficients different from 
zero in (\ref{rho}) increases, the approximation that 
leads to (\ref{rpert}) is not valid any more as the off-diagonal coefficients 
of (\ref{rho}) start to be not negligible.  
The overall effect is that the concurrence drops exponentially with 
$\ell$, up to a critical value where it
drops exactly to zero. This happens when all the off diagonal term are 
on average equal. As $\ell \propto k^2 $, 
small changes in the Hamiltonian (\ref{hamil}) of the kind $k
\rightarrow k +\Delta k$ change
exponentially the pairwise entanglement present in the system.


\begin{figure} 
\centerline{\includegraphics[height=6cm]{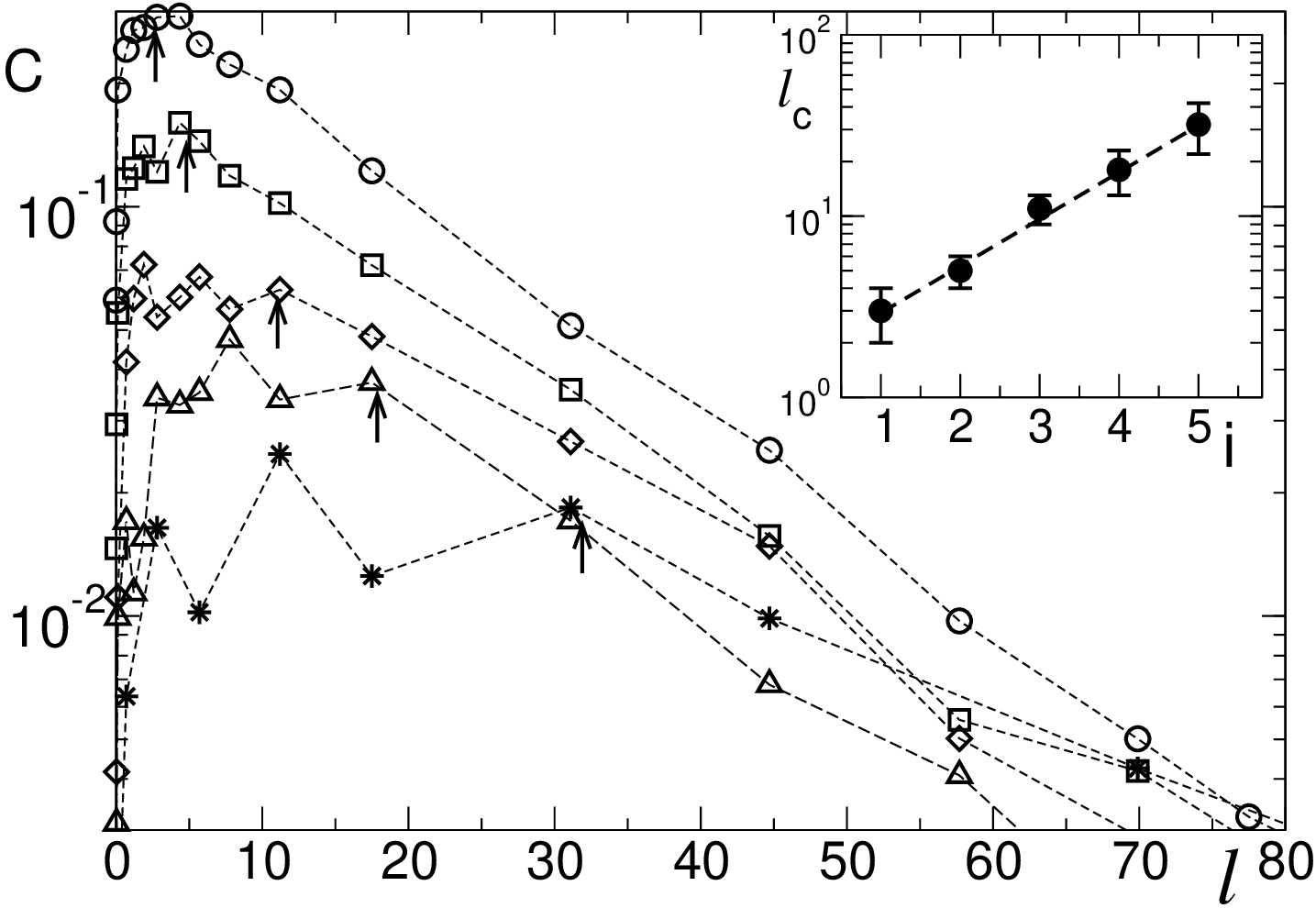}}
\caption{Concurrence saturation values between qubits $i$ and $j=i+1$
  as a function of localization length $\ell$. 
  Different simbols represents different qubits couples: $i=1$
  (circles), $i=2$ (squares), $i=3$ (diamonds), $i=4$ (triangles), 
  $i=5$ (stars).
  Arrows point where the exponential decay starts. 
  Inset: $\ell_c$ as a function of $i$. The dashed line is an
  exponential fit.}
\label{fig4}       
\end{figure} 
In the inset of Fig. \ref{fig3}  the 
concurrence saturation values for different qubit pairs are plotted
for a given value of the localization length.  As it can be clearly
seen the points follows the estimate $C(\ell) \propto 2^{-2j}$.
In Fig. \ref{fig4} we show that the two different regimes of the
concurrence are typical of any pair of the qubits:
For every couple of qubits chosen there is a regime of squared root 
increasing followed by an exponential decay. 
Notice that here the critical point $\ell_c$ where 
exponential sensitivitivy starts drastically depend on the couple of 
qubits chosen. Indeed, $\ell_c \sim 2^i$. This is due to the fact 
that the number of coefficients in (\ref{rho}) different from zero, 
scales as $\Delta \sim 2^i$, as the labels of the off-diagonal 
coefficients that are multiplied differ at least of $2^i$. 
Thus, not negligible off diagonal terms appears for greater values 
of $\ell$. In Fig. \ref{fig4} this behavior is shown by means of 
the arrows the predicted points $\ell_c(i)=2^i$: the value of the
localization length corresponding to the maximum concurrence 
saturation value are shown in the inset of Fig\ref{fig4}.


We now focus the bipartite entanglement and we characterizze it by
means of the Von Neumann entropy of one subsystem. We bipartite the
system in two subsystems $A$ and $B$, each composed by $n_A$ and $n_B$
qubits respectevely, we compute 
\begin{equation}
S_A = Tr \rho_A \mathrm{log} \rho_A,
\end{equation}
where $\rho_A$ is the reduced density matrix of the subsystem
$A$. Notice that, due to the hierarchy introduced by the binary coding
$n=\sum \alpha_i 2^i$, the bipartite entanglement displays very 
different behaviors depending on which qubits compose the
subsystems. It is necessary then to specify both the size and the
labels of the qubits in each subsystem. We first bipartite the system
in qubits one to $m$, varying $m$ ($n_A=m$ and $n_B=n_q-m$). 
Then, we study the case of subsystem $A$ composed of a single qubit
($n_A=1$ and $n_B=n_q-1$), varying its position. We evaluate the Von
Neumann entropy after the localization time, thus when the wave
function of the system can be described by Eq.(\ref{locpsi}) and $S_A$
is stationary. 

\begin{figure} 
\centerline{\includegraphics[height=6cm]{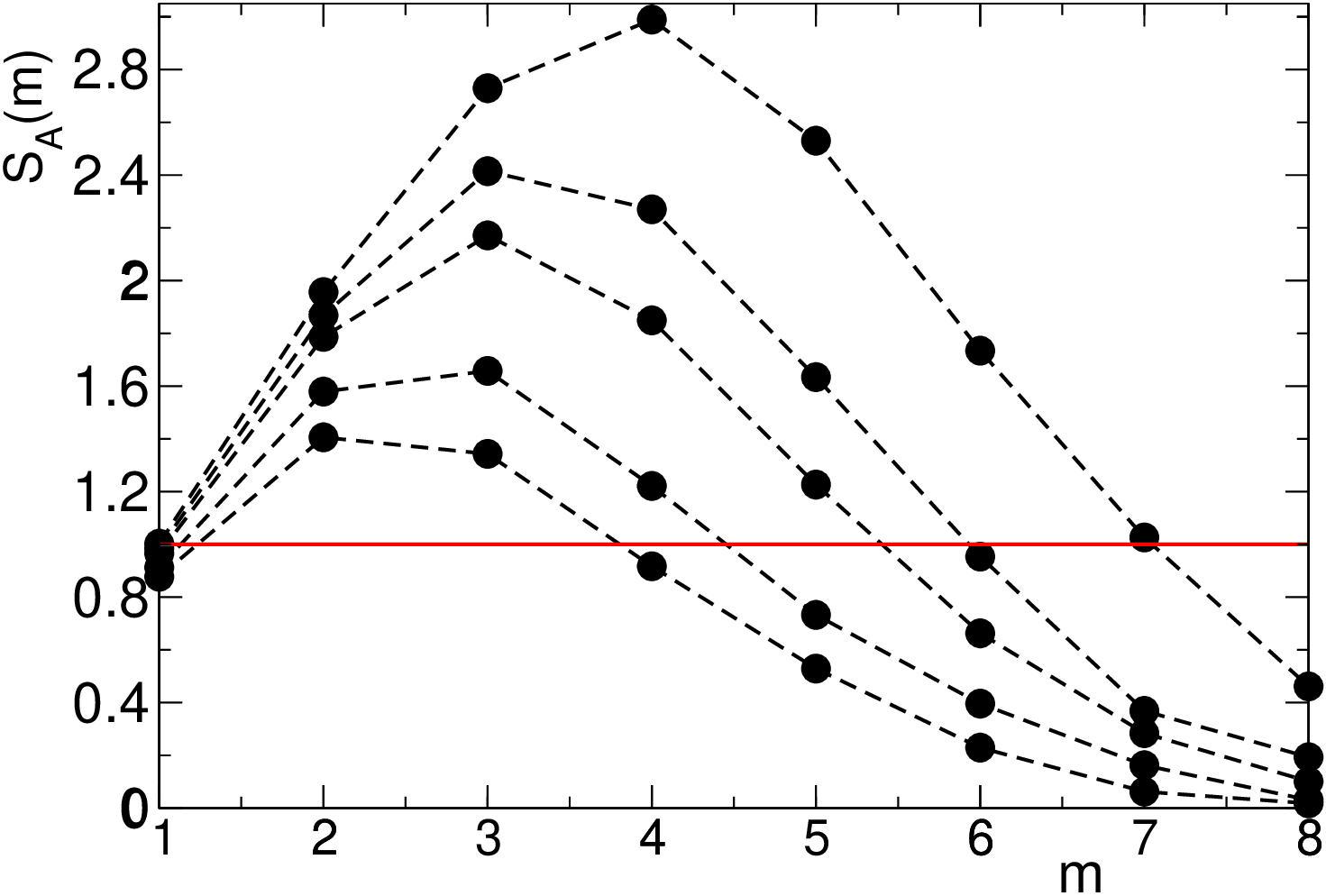}}
\caption{Von Neumann entropy $S_A$ of a subsystem composed by qubits
  $1$ to $m$ for different value of localization length (from bottom
  to top $\ell \simeq 2^k$, $k=4,5,6,7,8$). }
\label{fig5}       
\end{figure}

In Fig.\ref{fig5} we show the saturation level of the 
entropy of a subsystem of size $m$ for different values of the 
localization length $\ell$.  We define a critical threshold $S_c$ of the
entropy under which we consider the block of $m$ spins unentangled
with the others. If we choose $S_c=1$ (straight line in Fig.\ref{fig5}), it
is clear that the maximum number of qubits entangled increases as
$\mathrm{log}~ \ell$. It has been shown recently \cite{vidal} that the
feasibility of efficient classical simulations of a quantum system is 
conditioned by the presence of bipartite entanglement and its scaling
behavior with respect to the number of qubits $n$.
Indeed, if the maximum (with respect to the partition) bipartite
entanglement  in the system scales at maximun as $\mathrm{log}~ n_q$, it is
possible to perform efficient classical simulations. 
In our system, we may scale the number of qubits keeping fixed the 
system size, thus exploring smaller and smaller scales of the system 
($T \to 0$), or increasing the system size ($T=const$). In the former 
case $\ell$ grows exponentially with $n_q$ and $S_a \sim n_q$ thus
there are no known efficient classical simulation methods, while in the latter 
case $\ell$  is constant and it is thus possible to perform efficient 
classical simulations following \cite{vidal}.
This result should not surprise as increasing the system size
while studying a dynamically localized system adds almost no
information at all, as the wave function tail
are almost not populated (the wave function coefficients decays
exponentially). In conclusion, quantum computation of dynamically localized
systems do improve the efficiency of classical simulations. 
However, the improvement to compute interesting quantities as the
localization length is only quadratic and not exponential as shown
in \cite{proc}. The same arguments apply in the simulation of an
ergodic quantum chaotic system ($\ell=2^{n_q}$). Thus, the 
simulation of a complex many body quantum system, is classically 
inefficient while it is, at least in principle, possible to simulate
it efficiently if an efficient quantum algorithm exists. 

\begin{figure} 
\centerline{\includegraphics[height=6cm]{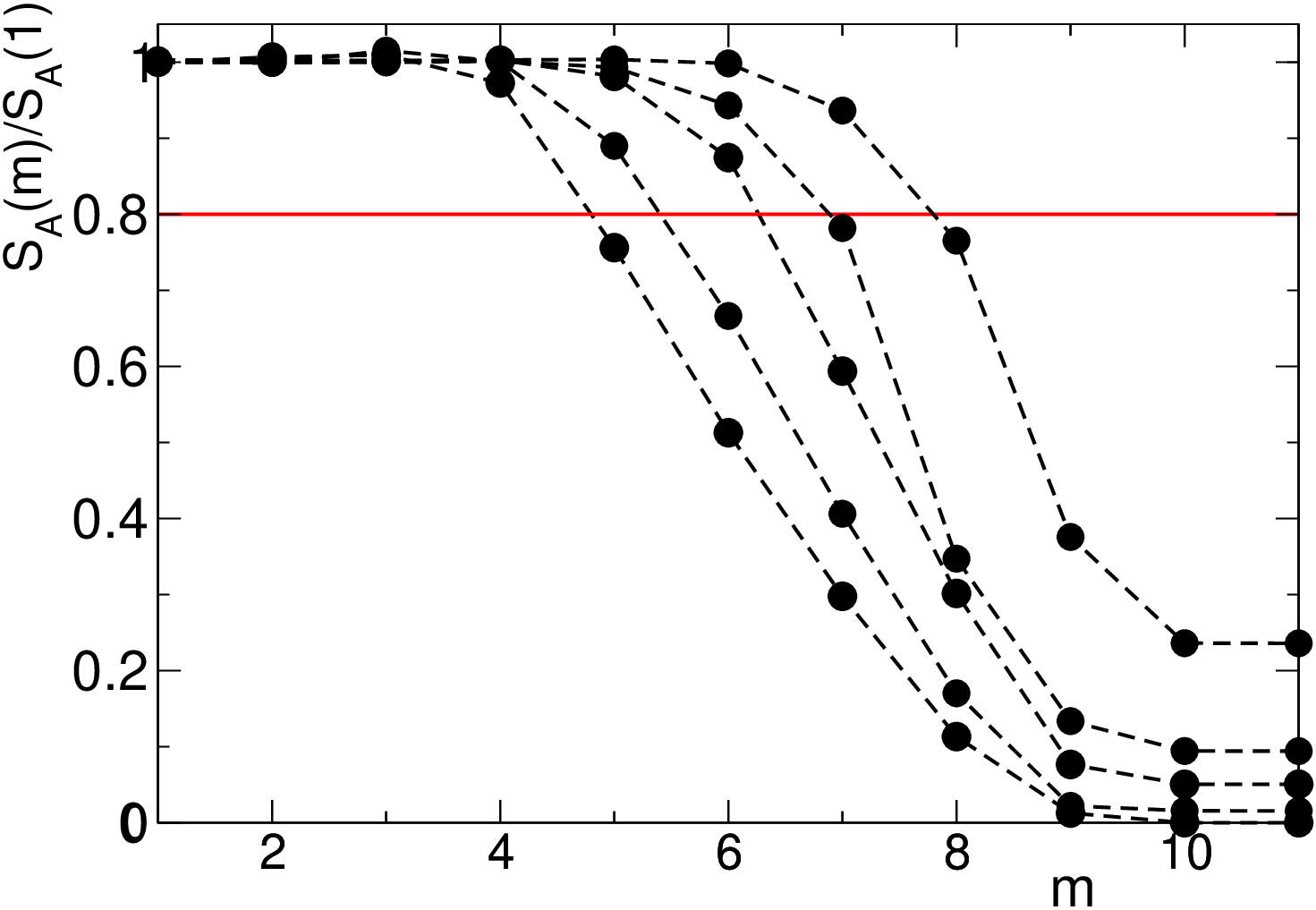}}
\caption{Von Neumann entropy $S_A$ of a subsystem composed by a single
  qubit $m$ for different value of localization length (from bottom
  to top $\ell \simeq 2^k$, $k=4,5,6,7,8$).}
\label{fig6}       
\end{figure}

In Fig. \ref{fig6} we perform a different analysis, studying the Von
Neumann entropy of a single qubit with respect to the rest of the
quantum computer as a function of the localization length. The figure
shows clearly that the entanglement depends on the position of the qubit.
Again, we define a critical threshold under which we consider not
entangled the qubit. As before, the number of entangled qubits scales
logarithmically with the localization length. Thus, the measure of the
reduced density matrix of a single qubit can be used as an efficient
method to estimate the localization length of a quantum chaotic
system. This estimate is crucial to investigate  
complex quantum system efficiently by means of both classical 
and quantum simulations.

In conclusion we have shown that the concurrence in a quantum computer 
that simulates a dynamically localized system is exponentially
sensitive to both small changes of the Hamiltonian and to the 
qubits chosen. This sensitivity is due to the natural ordering
introduced on the qubit by the coding of the simulated system.
Notice that this is a signature of quantum chaos on a pure quantum 
characteristic with no classical analogue. 
It should be also interesting to compare such sensitivity with the 
cases of classical regular and semi-integrable dynamics.
Furthermore, the same sensitivity has been found recently 
in the ground state of a single particle Anderson model  \cite{li}: 
these two results reflects the underlaying connections between 
the dynamical and the Anderson localization, and a better
comprehension of this behavior might lay in a more general picture. 
The results on the Von Neumann entropy showed  that, the dynamical
regime of the simulated quantum system influence 
the possibility of performing efficient classical simulations.
In the case of a localized state there are no known methods to 
perform an efficient classical simulation if one is exploring smaller scales
while increasing the number of qubits. 
A more detailed picture of the dependences of the
feasibility of efficient quantum simulations depending on the
dynamical regime of the system will be the objects of further studies.

The author thanks Rosario Fazio and Giuliano Benenti for interesting
discussions and a careful reading of the manuscript.

\end{document}